\documentclass[10pt,journal,compsoc]{IEEEtran}
\usepackage{color}

\ifCLASSOPTIONcompsoc
\usepackage[nocompress]{cite}
\else
\usepackage{cite}
\fi

\usepackage[utf8]{inputenc}
\usepackage{longtable}
\usepackage{graphicx}
\usepackage{color}

\usepackage[cmex10]{amsmath}

\usepackage{amsfonts}
\usepackage{bm}
\usepackage{color}
\usepackage{amsmath}
\usepackage{booktabs}
\usepackage{multirow}
\usepackage{soul}
\definecolor{dullYellow}{rgb}{0.8,0.8,0}

\usepackage{array}

\usepackage{mdwmath}
\usepackage{mdwtab}
\usepackage{subfig}
\usepackage{url}

\usepackage{multirow}
\usepackage{booktabs}
\usepackage{amssymb}
\usepackage{lscape}

\newcommand{\etal}{{\it{et al.}}}
\newcommand{\eg}{{\it{e.g., }}}
\newcommand{\ie }{{\it{i.e., }}}
\newcommand{\noi}{\noindent}

\newcommand{\ltwo}{\mathcal{L}_2}
\newcommand{\rtwo}{\mathbb{R}^2}

\newcommand{\branch}{\beta}		
\newcommand{\proot}{\bm{\beta}}	

\newcommand{\srvf}{q}
\newcommand{\srvft}{\bm{\srvf}}
\newcommand{\preshapetrees}	{\mathcal{C}}

\newcommand{\rotation}{O}
\newcommand{\reparam}{\gamma}

\newcommand{\meansrvft}{\bm{\mu}}   
\newcommand{\covMatrix}{C}          
\newcommand{\real}{\ensuremath{\mathbb{R}}}
\newcommand{\geod}{\bm{\alpha}}

\def\argmin{\mathop{\rm argmin}}

\begin{document}
\bstctlcite{IEEEexample:BSTcontrol}
%
%
%
\title{Statistical Analysis and Modeling \\of the Geometry and Topology of Plant Roots}

\author{Guan Wang, Hamid Laga, Jinyuan Jia, Stanley J. Miklavcic, Anuj Srivastava
	
	\IEEEcompsocitemizethanks{
	\IEEEcompsocthanksitem Guan Wang and Jinyuan Jia are with School of Software Engineering, Tongji University, China. E-mail: guan.wang@tongji.edu.cn
		\IEEEcompsocthanksitem  Hamid Laga  is with the Information Technology, Mathematics and Statistics Discipline, Murdoch University (Australia), and with the Phenomics and Bioinformatics Research Centre, University of South Australia.
		E-mail: H.Laga@murdoch.edu.au
		\IEEEcompsocthanksitem Stanley J. Miklavcic is with with the Phenomics and Bioinformatics Research Centre, University of South Australia.
		\IEEEcompsocthanksitem Anuj Srivastava is with Department of Statistics, Florida State University, USA. 
} 
	}
	
\markboth{Statistical Analysis and Modeling of the Geometry and Topology of Plant Roots}%
{Wang, Laga, Jia, Miklavcic and Srivastava \MakeLowercase{\etalnospace}: Statistical Analysis and Modeling of the Geometry and Topology of Plant Roots}

\IEEEcompsoctitleabstractindextext{
\begin{abstract}
The root  is an important organ of a plant since it is responsible for water and nutrient uptake. Analyzing and modelling variabilities in the geometry and topology of  roots can help in assessing the plant's health, understanding its growth patterns, and modeling relations between plant species and between plants and their environment. In this article, we develop a framework for the statistical analysis and modeling of the geometry and topology of  plant roots. We represent root structures as points in a tree-shape space equipped with a metric that quantifies geometric and topological differences between pairs of roots.  We  then use these building blocks to  compute geodesics, \ie optimal deformations under the metric  between root structures, and to  perform statistical analysis on root populations. We demonstrate the utility of the proposed framework through an application to a dataset of wheat roots grown in different environmental conditions.  We  also show that the framework can be used in various applications including classification and regression.
\end{abstract}
}


\maketitle


\IEEEdisplaynotcompsoctitleabstractindextext

%

\section{Introduction}
Roots are the primary site of nutrient and water uptake for plants and play a crucial role in plant growth. Understanding structural (\ie  geometric and topological) similarities and dissimilarities between roots, and capturing and modeling the structural variability in root populations  can help in  assessing the plant's health, understanding its interaction with the surrounding environment, and modelling growth processes.

%

Existing techniques for modelling structural variability are limited to objects that only bend and stretch~\cite{blanz1999A,kurtek2013landmark,laga2017numerical,laga2014landmark,jermyn2017elastic}. Early works, \eg~\cite{blanz1999A}, use morphable models, which represent the shape of an object using a dense set of landmarks and thus it can be seen as a point in a high-dimensional Euclidean space.  Statistical analysis can then be performed using standard techniques such as principal component analysis~\cite{laga20183d}. However, these techniques are  limited to objects which only slightly bend or stretch. Recent works such as~\cite{kurtek2013landmark,laga2017numerical,laga2012riemannian,laga2014landmark} proposed new formulations that are suitable for large elastic deformations.   These techniques, however, do not handle topological variations such as those encountered  in natural objects that have a tree structure\footnote{By a tree structure or a tree-shape, we refer to the mathematical concept of tree, \ie a graph of nodes and edges with no cycles. Each edge can have attribute that define its shape. As such, although a botanical tree and a root are two different things, mathematically they have a tree-shape or a tree structure. 
}, \eg  plant's shoots or roots, airway trees, and neuronal  structures in the human brain. In these types of objects, growth processes, disease progression, and environmental effects affect not only their geometry  but also their branching structures.


In this article, we propose a framework for the  statistical analysis  of the geometry and topology of  objects that have a tree structure. We focus on plant roots characterised by a main root and first order lateral roots. Higher order lateral roots (sub laterals and off laterals) are not catered for in the present form, although the model can be extended to treat these as well. In the meantime, we foresee that actual and more complex root systems are dissected  to attain the form applicable to the analysis presented here. The  framework we propose builds on the representation suggested by Duncan \etal~\cite{duncan2018statistical} for the analysis of simple neuronal structures. It  allows one to:
\begin{itemize}
  \item Compute correspondences and geodesic  paths between  plant roots  even when the latter undergo large bending, stretching, and topological deformations.

  \item Compute statistical atlases, \ie  the means and  principal modes of variation, of plant root collections.

  \item  Characterize the geometric and structural variability within a collection of roots using   probability distributions.

  \item Develop a mechanism a program in which plant roots are synthesized either randomly, using random sampling, or in a controlled manner using regression from biologically motivated parameters. 

\end{itemize}
\noi  The proposed framework has a wide range of applications. We show its utility in root classification and root synthesis. It also has multiple applications in plant biology including  (1) quantifying differences in root morphology, and (2) comparing root systems of either different genotypes grown under the same environmental conditions, or root systems of the same genotype plant grown under different environmental conditions, including different nutrient concentrations, or soils at different levels of moisture content or different toxicity.


The remaining parts of this article are organised as follows; Section~\ref{sec:related_work} overviews the related work. Section~\ref{sec:math} lays down the mathematical formulation of the concepts tree-shape space and its associated metric and geodesics. These   concepts  are then used to perform statistical analysis of collections of plant roots (Section~\ref{sec:statistical_atlas}), and to synthesize and simulate 3D plant roots (Section~\ref{sec:synthesis}).  Section~\ref{sec:results} presents the results, and compares the performance of the proposed approach and the quality of its  results with the state-of-the-art. Section~\ref{sec:conclusion} summarizes the main findings of this article.

%

\section{Related works}
\label{sec:related_work}

The framework developed in this article can be seen as the generalization to objects with tree-like structures (\eg plant roots and shoots) of the statistical analysis techniques that have been proposed for manifold 3D shapes.  These types of objects vary not only in geometry but also in topology. Thus, we focus our survey on the techniques which have been proposed for capturing root morphology, and on techniques for statistical shape analysis.

\subsection{Root morphology analysis}

Quantitative characterization of root shapes can reveal fine differences between various phenotypes, signals, and exogenous regulatory substances~\cite{grabov2005morphometic}. This can facilitate molecular-genetic and physiological studies of root development. Some approaches extract shape descriptors to capture the important geometric and morphological properties of root shapes. Examples of such properties include  the angular deviation of the root tip from vertical axis~\cite{marchant1999aux1,fujita1997pisl,fujita1996genetic}, the vertical growth index~\cite{grabov2005morphometic,Vaughn2011AQTL}, which is measured as the ratio between the root tip ordinate and the  root length, and the horizontal growth index defined as the ratio between the root tip abscissa and the root length.  Schultz \etal~\cite{Schultz2016root} employed straightness, wave density, and horizontal growth index to describe the root shape.

These shape descriptors have been used to represent and compare root shapes in the descriptor space. They, however,  only represent spatial information about roots. Also,  variability in the descriptor space often does not correspond to  variability in root shapes. For example, the average of two root descriptors is  not guaranteed to correspond to the average root, not even to a valid root shape.

\subsection{Statistical Shape analysis and modeling}

Instead of using  descriptors to characterize the shape of objects,  several recent papers treat the shape of an object as a point in a shape space. Equipping the shape space with a proper metric allows for a comparison of objects based on their shapes, computing geodesics, and performing statistical analysis including regressions and shape synthesis. These concepts have been extensively investigated  in the case of 3D manifold objects such as human faces~\cite{blanz1999A}, human bodies~\cite{allen2003the}, arbitrary natural objects~\cite{kurtek2013landmark,laga2017numerical}, and man-made objects~\cite{laga2017modeling}.  These methods, however, are limited to 3D models with fixed topology. They cannot capture and model structural variabilities such as those present in plant roots.

More closely related to our approach are techniques based on tree statistics. The seminal work of Billera \etal~\cite{billera2011geometry}   proposed the notion of continuous tree-space and its associated tools for computing summary statistics. Some variants of this idea have been introduced for the statistical analysis of tree-structured data, \eg~\cite{owen2011A,aydn2009a}. These works, however,  only consider the topological structure of trees and ignore the geometric attributes of edges, which limits their usage. To overcome this limitation, several extended methods have been proposed for defining a more general tree-space. This includes Feragen \etal's framework~\cite{feragen2010geometries,feragen2011means,feragen2013toward,feragen2013tree-shapce}, which proposed a tree-shape space for computing statistics of airway trees, and its extension to complex botanical trees~\cite{wang2018the,wang2018statistical}.

Despite their efficiency and accuracy in certain situations, these techniques exhibit three main fundamental limitations. First, they use the Quotient Euclidean Distance (QED), which is not suitable for capturing large elastic deformation, \ie bending and stretching, of the branches. Second, they represent tree shapes as \emph{father-child branching} structure, which leads to a significant shrinkage along the geodesics between trees that exhibit large topological differences. Third,  branch-wise correspondences  need to be manually specified, especially when dealing with complex tree-like structures,  which restricts  its utility in practical applications.

\vspace{6pt}
\noi \textbf{Contributions: } We propose in this article a statistical framework that is more suitable for analyzing plant root shapes. It  builds upon and extends  the recent work of Duncan \etal~\cite{duncan2018statistical}. First, we show that the proposed \emph{main-side branching representation} is more efficient for capturing topological changes than the \emph{father-child branching} employed in the frameworks of Feragen \etal~\cite{feragen2010geometries,feragen2011means,feragen2013toward,feragen2013tree-shapce} and Wang \etal~\cite{wang2018the,wang2018statistical}. Second, we use the new representation to develop tools for computing, jointly,  one-to-one correspondences and geodesic deformations between plant roots that significantly differ in geometry and topology. Finally, we demonstrate the utility of the framework in high-level applications such computing statistical summaries and atlases, synthesising root shapes, either randomly or in a controlled manner, and classifying plant roots based on their geometry and topology. To the best of our knowledge, this is the first approach that deals with statistical modeling of plants root based on their shapes.

\section{Mathematical representation and shape space metric}
\label{sec:math}
\begin{figure}[t]
\center
\includegraphics[width=0.3\textwidth]{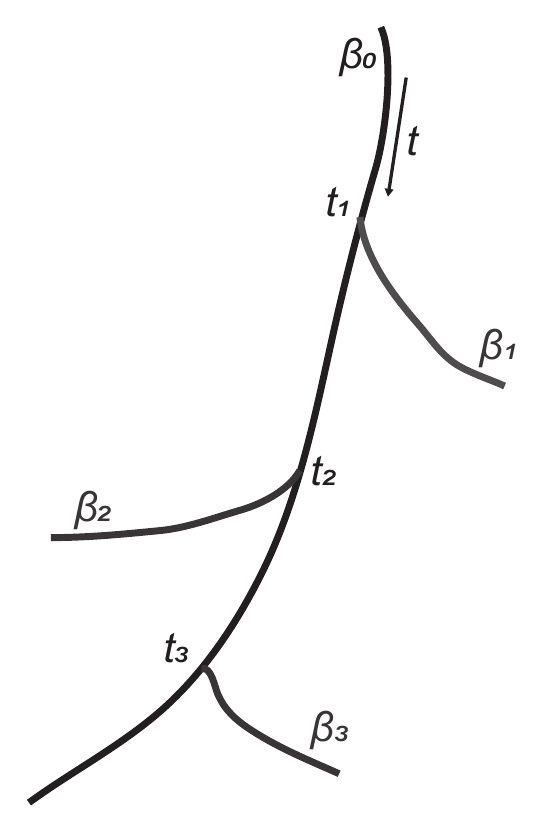}
\caption{The representation of a simple plant root.}
\label{fig:simple_root_representation}
\end{figure}

The input to our framework is a collection of 2D plant roots. We first skeletonize each root and convert it into a set of  two-layer planar curves: ${\proot} = (\branch_0,\{\branch_i\}_{i=1}^n)$. Here, $\branch_0$ represents the main root branch and $\{\branch_i\}_{i=1}^n$ represent  the finite (possibly empty) set of lateral branches, each one is attached to the main root  at some locations, see Figure~\ref{fig:simple_root_representation}. Each branch $\beta_i$ is a continuous curve in $\rtwo$, \ie $\beta_i\ : [0, 1]\rightarrow \rtwo$. Let $t_i \in [0, 1]$  such  that $\beta_0(t_i) = \beta_i(0)$ is the location  on $\beta_0$ to which the $i-$th lateral branch $\beta_i$ is attached.



In order to perform statistical analysis of the geometry and topology of a collection of such roots, we need to define a tree-shape space\footnote{A tree is a graph with no cycles. It is defined as a set of nodes and edges that connect them. Each edge can also be augmented with shape attributes. },  a metric on this space, and a mechanism for computing correspondences and geodesics in the tree-shape space.    Due to their arbitrary structure, two roots will rarely have the same number of lateral roots. To simplify the search for correspondences, and following~\cite{duncan2018statistical}, we augment each root $\proot$   by adding virtual lateral branches, \ie branches of length zero,  according to the $t$ value of each non-virtual lateral branch. For example, for two roots ${\proot_1} = (\branch_0^1,\{\branch_i^1\}_{i=1}^{n_1})$ and ${\proot_2} = (\branch_0^2,\{\branch_i^2\}_{i=1}^{n_2})$, we add $n_2$ null branches to $\proot_1$, whose locations on $\branch_0^1$ are computed as $t^1_{n_1+i} = t^2_i, i=1,2,\dots,n_2$. Then, we add $n_1$ null branches to $\proot_2$ in the same manner.
In what follows, and for simplicity of notation, we will also use the symbol $\proot$ to indicate an augmented root.



We represent each root $\proot$ using its Square Root Velocity Function Tree (SRVFT)~\cite{srivastava2011shape}, $\srvft$, which consists of a collection of  Square Root Velocity Functions, one for each branch $\branch$ in $\proot$.  That is, given one branch $\branch$, its SRVF representation is defined as its derivative scaled by the square-root of the $\ltwo$ norm of the derivative:
\begin{eqnarray}
\label{equ:SRVF_representation}
\srvf(t) =
\begin{cases}
\frac{\dot{\branch}(t)}{\sqrt{\parallel\dot{\branch}(t)\parallel}}       & \text{ if } \dot{\branch}(t)  \text{ exists and is nonzero,} \\
0   & \text{otherwise.}
\end{cases}
\end{eqnarray}

\noi As such, the SRVF is translation invariant. Thus, when converting an entire root into its SRVFT representation, we lose track of the location of the bifurcation points.  In order to preserve them,   we represent each side branch using an ordered pair $(\srvf_i, s_i)$ where $s_i \in [0, 1]$ is the starting location on $\beta_0$. Then, the SRVFT representation of the entire root $\proot$ becomes $\srvft = (\srvf_0, \{(\srvf_i, s_i)\}_{i=1}^N)$.

Let $\preshapetrees$ denote the space of all root shapes, which are represented by their SRVFTs.  $\preshapetrees$ is called \emph{pre-shape space}~\cite{srivastava2011shape}. We now need to define the metric and the mechanism for computing geodesics between a pair of roots.


\subsection{Metric for tree-like shapes}

Our goal is to define a metric, which quantifies the geometric, \ie bending and stretching,  and topological deformations. We follow the approach of Duncan \etal~\cite{duncan2018statistical}. Let $\srvft^i = (\srvf_0^i, \{(\srvf_k^i, s_k^i)\}_{k=1}^N), i \in \{1, 2\}$ be  the SRVFT representations of two roots $\proot_1$ and $\proot_2$, respectively.   We define the distance, or dissimilarity, between such roots in $\preshapetrees$ as:
\begin{equation}
\scriptsize
\label{equ:init_distanceTwoRoots}
	 d_\preshapetrees(\srvft^1, \srvft^2) = \lambda_m \parallel\srvf_0^1 - \srvf_0^2\parallel^2 + \lambda_s \sum_{k=1}^N \parallel\srvf_k^1 - \srvf_k^2\parallel^2 + \lambda_p \sum_{k=1}^N ({s}_k^1 - {s}_k^2)^2,
\end{equation}

\noi which is a weighted sum of three terms. The first term is the $\ltwo$ distance between the main roots. Since the $\ltwo$ distance in the SRVF space is equivalent to the full elastic metric in the space of curves~\cite{srivastava2011shape}, then the first term measures the amount of bending and stretching needed to align one curve onto the other. The second term is the $\ltwo$ distance between the SRVFs of the lateral roots. The last term measures the distance between the  bifurcation points and is used to capture topological changes. The parameters $\bm{\lambda} = (\lambda_m, \lambda_s, \lambda_p)$ control the relative contributions of the three terms.


A good metric for statistical shape analysis should be invariant to shape-preserving transformations, \ie translation, scaling, rotation, and re-parameterization. The SRVF representation is translation-invariant by construction since it uses  derivatives. Invariance to scale can be efficiently handled by scaling each root by the length of its main root. Note that the latter might not be needed depending on the application at hand. For instance, growth analysis requires preserving the scales of the roots.

The remaining variabilities, \ie rotation and reparameterization, are handled algebraically following~\cite{laga2014landmark,duncan2018statistical}.
The action of all possible rotations  $\rotation \in SO(2)$  and reparameterizations  $\reparam \in \Gamma$ on a root shape $\proot$ forms a set of roots  that have the same  shape ($\Gamma$ here is the space of all orientation-preserving diffeomorphisms of $[0, 1]$ to itself). Similar to~\cite{laga2014landmark}, we  redefine the dissimilarity between two roots $\proot_1$ and $\proot_2$ as the length of the shortest geodesic between $\srvft^1$ and $O(\srvft^2, \gamma)$ (here,  $(\srvft, \gamma)$ is the SRVFT of $\proot \circ \gamma$):
\begin{equation}
\label{equ:distanceTwoRoots}
	d({\proot_1}, {\proot_2}) = \min_{\rotation \in SO(2), \gamma \in \Gamma} d_\preshapetrees(\srvft^1, O(\srvft^2, \reparam)),
\end{equation}

\noi where $d_\preshapetrees(\srvft^1, O(\srvft^2, \gamma))$ is the length of the geodesic between $\srvft^1$ and the rotated and re-parameterized version of $\srvft^2$. The optimization over $O$ and $\gamma$ is the registration process, which consists of searching for  optimal alignment and correspondence between $\proot_1$ and $\proot_2$. It is solved as a linear assignment problem following~\cite{duncan2018statistical}.

\subsection{Geodesics}

A geodesic is the optimal deformation (bending, stretching, and topological changes), under the metric,  from one root to another. Its length quantifies the minimum amount of deformation that one needs to apply to one root in order to align it onto the other.   For simplicity, let us also denote by $\srvft^2$   the version  of $\srvft^2$ after applying the optimal rotation and reparameterization found by optimizing Equation \eqref{equ:distanceTwoRoots}.  Since the distance between $\srvft^1$ and $\srvft^2$ is a weighted sum of Euclidean distances, see Equation~\eqref{equ:init_distanceTwoRoots}, then the geodesic $\geod$ connecting them is the linear path that connects the two points, \ie:
\begin{equation}
	\label{equ:geodesicComputation}
    	\geod(r) = (1-r)\srvft^1 + r\srvft^2, 0 \leq r \leq 1,
\end{equation}

\noi which is defined in the SVFT space. The geodesic between   $\proot_1$ and $\proot_2$ is then given by mapping the path $\geod$ back to the space of trees.

\begin{figure}[t]
\center
\includegraphics[width=0.5\textwidth]{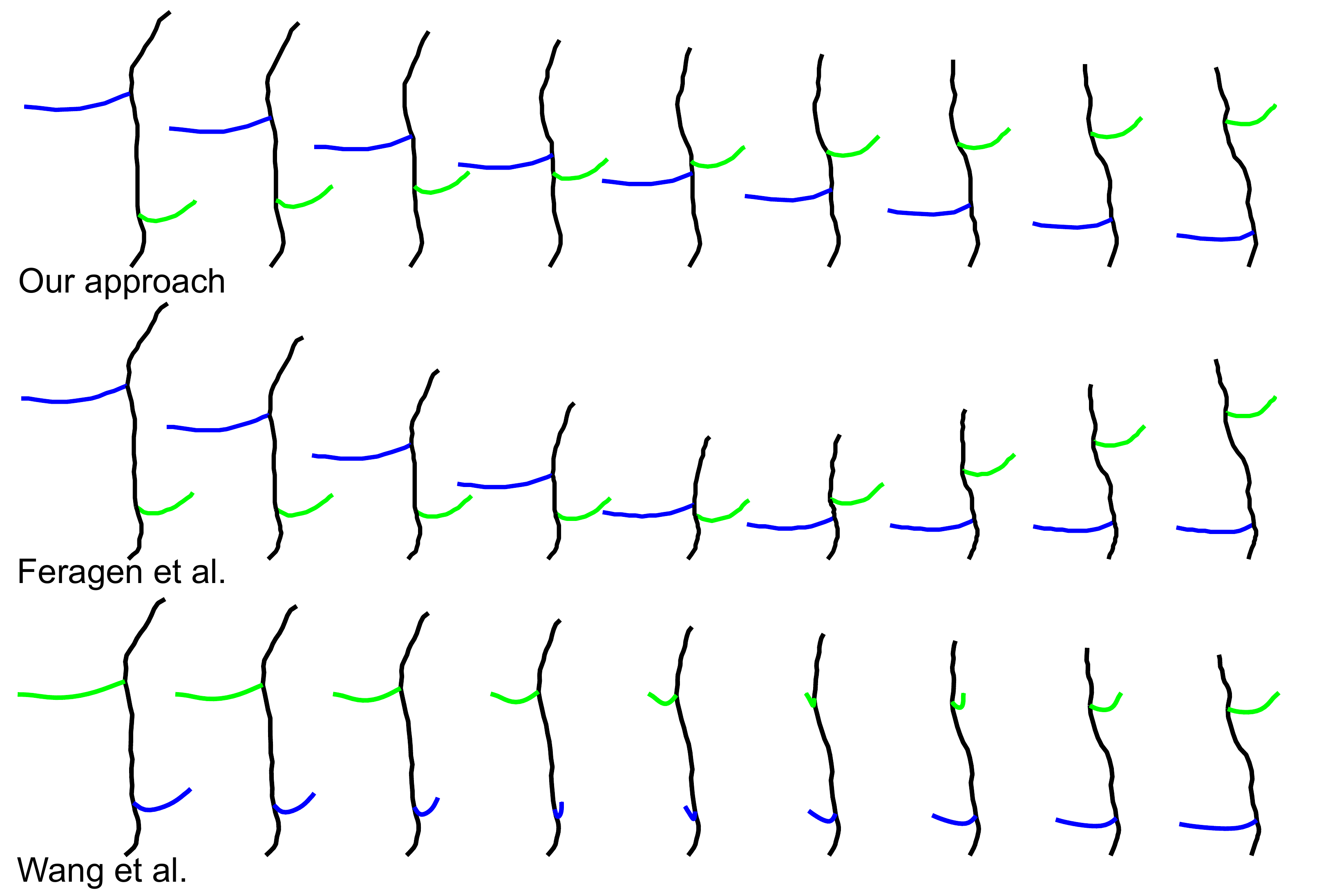}
\caption{Comparison between the geodesics obtained using our method (top row, $\lambda_m$ = 0.02, $\lambda_s $=1.0, $\lambda_p$ = 1.0), the approach of Feragen \etal~\cite{feragen2013toward} (middle row), and the approach of Wang \etal~\cite{wang2018the}  (bottom row).  The colors indicate branchwise correspondences. }
\label{fig:geodesicComparison_simple}
\end{figure}

Figure~\ref{fig:geodesicComparison_simple} shows an example of the geodesic between two simple tree-like shapes that differ in topology. For this example, we also show the results obtained by the approach of Feragen \etal~\cite{feragen2013toward} and the approach of Wang \etal~\cite{wang2018the}. Each row in the figure is one geodesic between the leftmost and rightmost roots. Branchwise correspondences are indicated using colors.   In this example, we can see that the intermediate shapes  along the geodesic obtained using~\cite{feragen2013toward}  are not natural; they exhibit unnatural shrinkage and are unable to find the correct correspondences.  The approach of Wang \etal~\cite{wang2018the} is less prune to global shrinkage. However, as seen in the last row of Figure~\ref{fig:geodesicComparison_simple}, it can match branches located in different sides. As such, the lateral branches unnaturally shrink and expand along the geodesic. 


\section{Statistical atlas}
\label{sec:statistical_atlas}
For a given set of root samples, denoted as $\{\proot_i, i = 1,2,...,m\}$, the mean root is defined as the root that is as close as possible to all the roots in the set. Mathematically, it is the root that minimizes the sum of the geodesic distances to all the  roots. Similar to geodesic computation, we first add to every root  $\proot_i$ virtual lateral branches to equalize the number of lateral branches across the roots in the dataset.
Let $\meansrvft$ be the SRVFT representation of the mean root. It can be computed as:
\begin{equation}
	\meansrvft = \argmin_{\srvft\in \mathcal{S},\rotation\in SO(2), \gamma \in \Gamma} d_\preshapetrees(\srvft, \rotation(\srvft^i, \gamma)).
	\label{eq:karcher_mean}
\end{equation}

\noi Here, $\mathcal{S}$ is the tree-shape space.  The solution to Equation~\eqref{eq:karcher_mean} is known as the Karcher mean. We employ the same gradient descent approach described in~\cite{laga2014landmark} to solve this optimization problem.

\begin{figure*}[t]
\center
\includegraphics[width=1.0\textwidth]{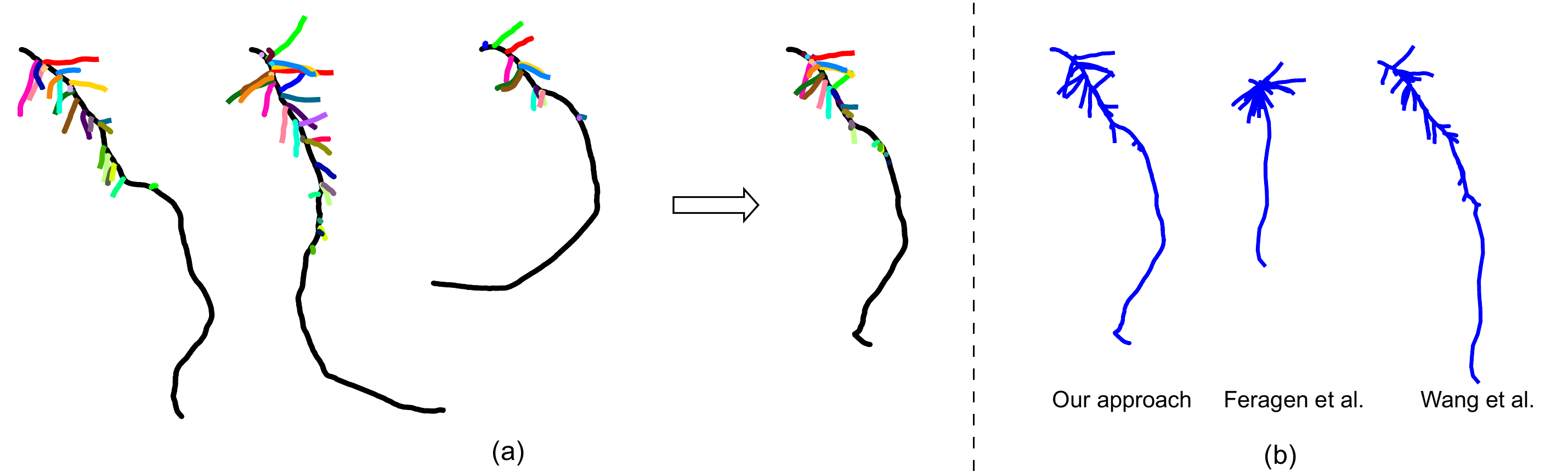}
\caption{(a) Mean root of three plant roots. Colors indicate branchwise correspondences.  (b) Comparison with Feragen \etal~\cite{feragen2013toward}  and Wang \etal~\cite{wang2018the}.}
\label{fig:meanRootResult_withComp}
\end{figure*}

Figure~\ref{fig:meanRootResult_withComp}(a) shows the mean root of three roots generated by this approach. Figure~\ref{fig:meanRootResult_withComp}(b) compares between the mean roots generated by our approach, the approach of Feragen \etal~\cite{feragen2013toward}, and the approach of Wang \etal~\cite{wang2018the}. It can be clearly seen that the mean root generated by our approach is more plausible than those generated by the other two methods.

In addition to the mean root, which can be regarded as a template that characterizes the main morphological properties of the shape of roots in the dataset,  one would also like to (1) quantify how the samples in the dataset deviate from the mean root, and (2)  analyze the distribution of roots   around the mean root. This can be done by computing the covariance matrix and the modes of variation. The tree-shape space $\mathcal{S}$ is a non-linear manifold, thus we employ its tangent space $T_{\meansrvft}(\mathcal{S})$ instead of $\mathcal{S}$ to perform statistical analysis. Similar to~\cite{srivastava2011shape,laga2014landmark}, we first map each $\srvft^i$ onto the tangent space using the inverse exponential map: $\srvft^i\rightarrow \bm{v^i} = \log_{\meansrvft}(\srvft^i)$. Then, we calculate the covariance matrix:
$\covMatrix = \frac{1}{m-1}\sum_{i=1}^m\bm{v^i}\bm{(v^i)^t}$. After calculating the eigenvalues $\{\lambda_i\}$ and eigenvectors $\{\Lambda_i\}$ of $\covMatrix$, we can obtain the principal modes of shape variation for the root collection. We also can generate a sample $v_k \in T_{\meansrvft}(\mathcal{S})$  along each eigenvector $\Lambda_i$ as: $v_k = \alpha\sqrt{\lambda_i}\Lambda_i, \alpha\in\real$ on the tangent space, where $\lambda_i$ is the eigenvalue associated with the eigenvector $\Lambda_i$.

\section{Plant roots  synthesis}
\label{sec:synthesis}
\subsection{Root synthesis by random sampling} 
The eigenvectors $\{\Lambda_i\}$ form an orthonormal basis of a Euclidean vector space. One can characterize the distribution of the input roots by fitting probability models either from a non-parametric family by computing their probability densities directly from the data, or from a parametric family, such as multivariate Gaussians. In this article, for the sake of simplicity, we fit to the population a multivariate Gaussian with mean $\meansrvft$ and a diagonal covariance matrix $\covMatrix$ whose diagonal elements are the square roots of the eigenvalues $\lambda_i$.

Once we have the probability model, a new botanical tree can be generated by randomly sampling from the distribution. In the case of a Gaussian model, a random root can be synthesized by first randomly generating $m$ real values $b_1,b_2, ..., b_n \in \real$ and then setting
\begin{equation}
	\srvft = \text{Exp}_{\meansrvft}\left(\sum_{i=1}^m b_i \sqrt{\lambda_i} \Lambda_i\right).
	\label{eq:pca_param}
\end{equation}

\noi Here,  $\text{Exp}_{\meansrvft}$ is the exponential map, which maps elements in the tangent space to $\mathcal{S}$ at $\meansrvft$ to the SRVFT space. A root $x$ can be obtained by mapping $\srvft$ back to the tree-shape space. In this article, we only consider the $m$-leading eigenvalues such that $\frac{\sum_{i=1}^{n}  \lambda_i}{\sum_i \lambda}>0.99$.  In order to generate plausible roots, one can restrict $b_i$'s to be within a certain range, \eg $[-1, 1]$.

\subsection{User-controlled root synthesis}

While random sampling from the multivariate Gaussian distribution provides an easy way of synthesizing new roots, it lacks control. In fact, sometimes, users would like to generate roots  by just adjusting a few parameters, \eg biologically motivated parameters.   In this article, we formulate this problem as a regression in the tree-shape space\textcolor{dullYellow}{~\cite{wang2018statistical}}. In particular, let $\textbf{p} \in \real^l$ be the vector of  $l$ parameters. Let   $x$ be a point in the tree-shape space and $\srvft$  its SRVFT  representation.  From Equation~\eqref{eq:pca_param}, $\srvft$ can be represented as a real valued vector $\textbf{b} = (b_1, b_2, \dots, b_m)^\top$.  If we assume that the relation between the biological parameters $\textbf{p}$ and the root to be linear, then the mapping can be represented using $m \times (l+1)$ matrix $\textbf{M}$ such that:
\begin{equation}
	\textbf{M}[p_1, p_2, \dots, p_l, 1]^\top = \textbf{b}.
\end{equation}

\noi After assembling all the parameter vectors into an $(l+1)\times m$ matrix $\textbf{P}$ and all the vectors $\textbf{b}$  into an $n \times m$ matrix \textbf{B}, the mapping matrix $\textbf{M}$ can be calculated as follows;
\begin{equation}
	\textbf{M} = \textbf{B}\textbf{P}^+,
\end{equation}

\noi where $\textbf{P}^+$ is the pseudoinverse of $\textbf{P}$.

There are several biologically-motivated parameters that can be used. In this article, we consider  (1) the main root length, (2) the mean length of the side-roots, and (3) the standard deviation of the side root lengths. We  first extract  these three parameters from a training dataset  and use them to learn the regression model. At runtime, the user can specify these parameters and the system automatically synthesizes new root models.

\section{Results and discussion}
\label{sec:results}
In this section, we consider a collection of wheat roots that exhibit different degrees of geometric and structural variations, and demonstrate the results of the proposed approach in computing (1) geodesics and (2) statistical summaries such as means and modes of variations. We also report the timing and compare our results to the state-of-the-art. Additional results are included in the supplementary material. For this, we have collected $53$ wheat plant roots with rich structural variations. First, wheat plants have been taken from soil, washed, and scanned using a flatbed scanner. Their images have then been binarized  and automatically skeletonized and converted into a two-layer structure representation as described in Section~\ref{sec:math}.

\subsection{Geodesics}

\begin{figure*}
\center
\includegraphics[width=1.0\textwidth]{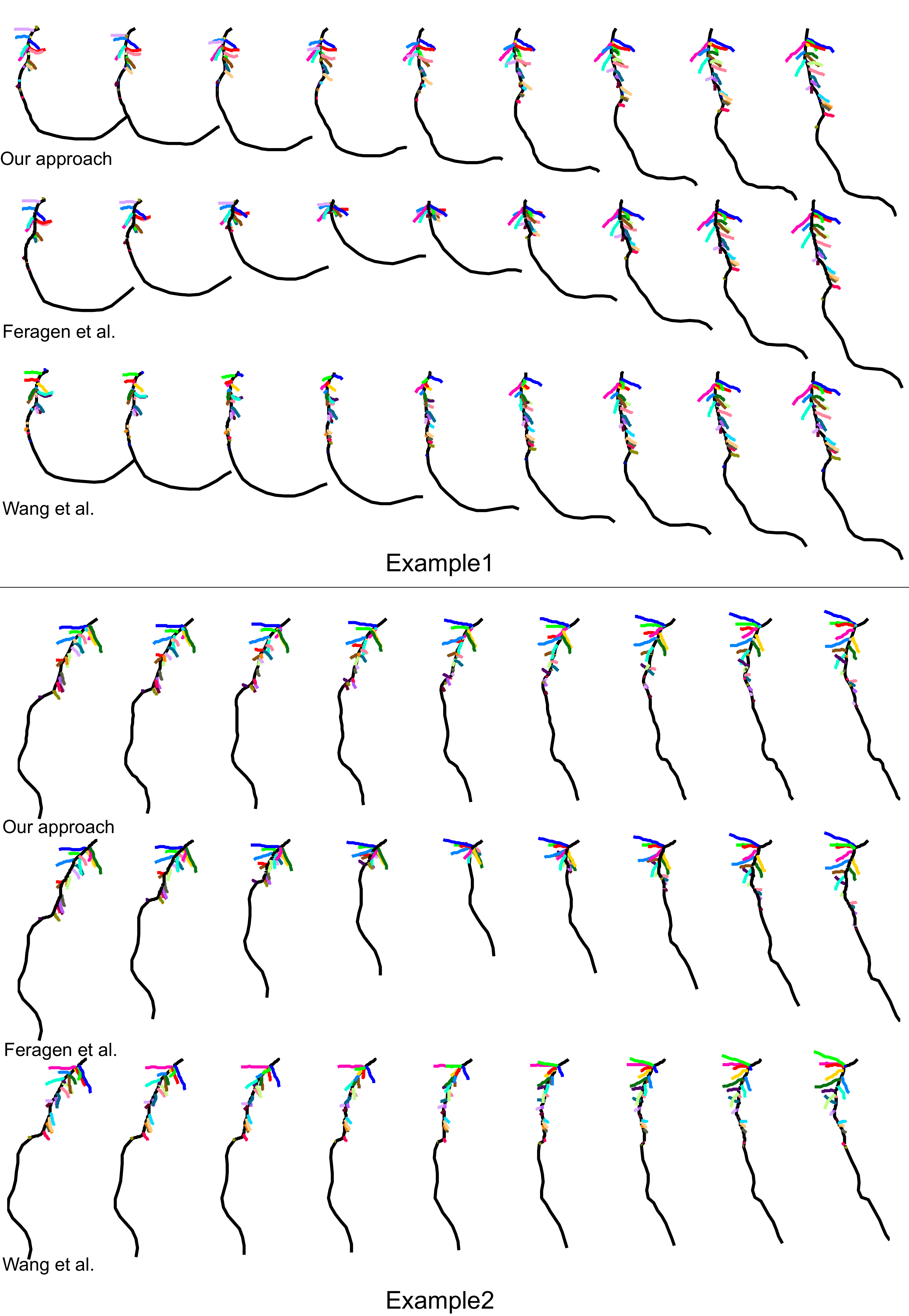}
\caption{Examples of geodesics between the most left and the most right plant root. For each example, we show the geodesics obtained using the approach proposed in this article (top row, $\lambda_m$ = 0.02, $\lambda_s $=1.0, $\lambda_p$ = 1.0 in each example), the approach of Feragen \etal~\cite{feragen2013toward} (middle row), and the approach of Wang \etal~\cite{wang2018the} (bottom row). Colors indicate branch-wise correspondences. }
\label{fig:geodesicComparison_1}
\end{figure*}

In this experiment, we consider pairs of plant roots, each pair is composed of a source and a target root,  and generate new roots by computing the geodesic (optimal deformation) that connects the source to the target. Figure~\ref{fig:geodesicComparison_1} shows two examples of such geodesics. For comparison, we also show geodesics between the same pair of roots but computed using the approaches of Feragen \etal~\cite{feragen2013toward} and of Wang \etal~\cite{wang2018the}. In each example, the colors indicate the branch-wise correspondences.

As one can see, the geodesics generated with the approach of Feragen \etal~\cite{feragen2013toward} exhibit unnatural shrinkage: the intermediate roots along the geodesic shrink and then expand. Also, the approach of Wang \etal~\cite{wang2018the}  fails to find correct branch-wise correspondences, which significantly affects  the quality of the geodesics obtained with this approach. In comparison, the approach proposed in this article is able to produce smooth geodesics with plausible-looking intermediate roots. More results are included in the supplementary material.  

\begin{figure*}[t]
\center
\includegraphics[width=1.0\textwidth]{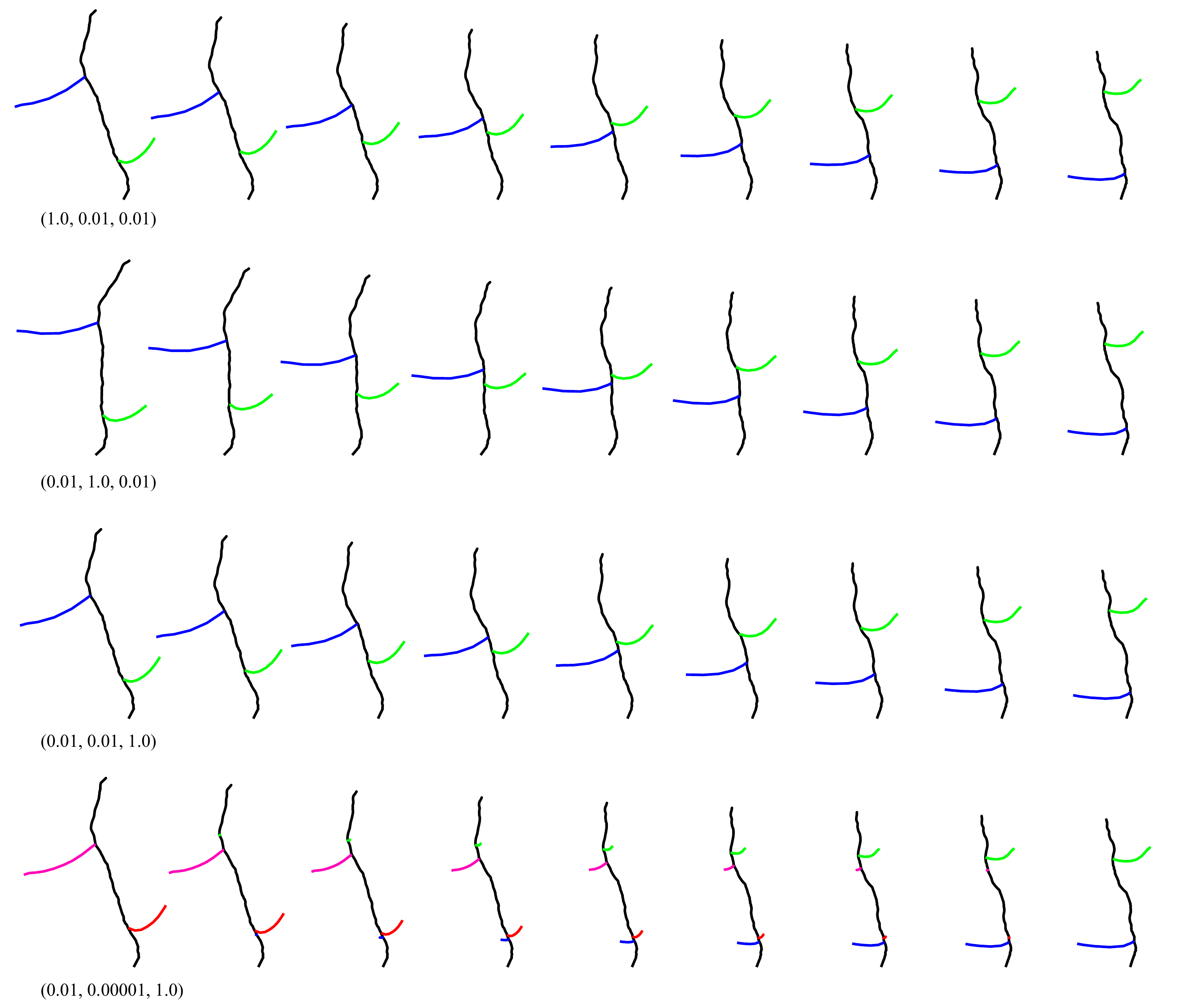}
\caption{The influence of weights of the three terms of  Equation~\ref{equ:init_distanceTwoRoots}  on the geodesics obtained with our appraoch. $(\lambda_m,  \lambda_s, \lambda_p) = (1.0, 0.01, 0.01)$ for the first row,   $(0.01, 1.0, 0.01)$  for the second row, $(0.01, 0.01, 1.0)$ for the third row, and $(0.01, 0.00001, 1.0)$ for the fourth row. Colors indicate branch-wise correspondences. }
\label{fig:geodesicComparison_parameters}
\end{figure*}

In addition, we check the influence of the weight $\lambda_m, \lambda_s$, and $\lambda_p$ of Equation~\eqref{equ:init_distanceTwoRoots}  on the quality of the geodesics. For the purpose of this experiment, we vary the values of these parameters and observe the results. This is illustrated in Figure \ref{fig:geodesicComparison_parameters} on a toy example.  In the first three results where the triplet $(\lambda_m,  \lambda_s, \lambda_p)$ is set to $(1.0, 0.01, 0.01), (0.01, 1.0, 0.01), (0.01, 0.01, 1.0)$, respectively,  the three geodesics look very similar. However, in the fourth result where $\lambda_m = 0.01$,  $\lambda_s = 0.00001$, and  $\lambda_p = 1.0$, the approach favours the creation of new lateral branches rather than sliding the existing one. This is predictable since the last term, which measures distance between bifurcation points, is highly penalized.    

The supplementary material also includes additional complex examples that show the effect of varying the parameters of Equation~\eqref{equ:init_distanceTwoRoots} on  the quality of the geodesics.


\begin{figure*}[t]
\center
	 \includegraphics[width=1.0\textwidth]{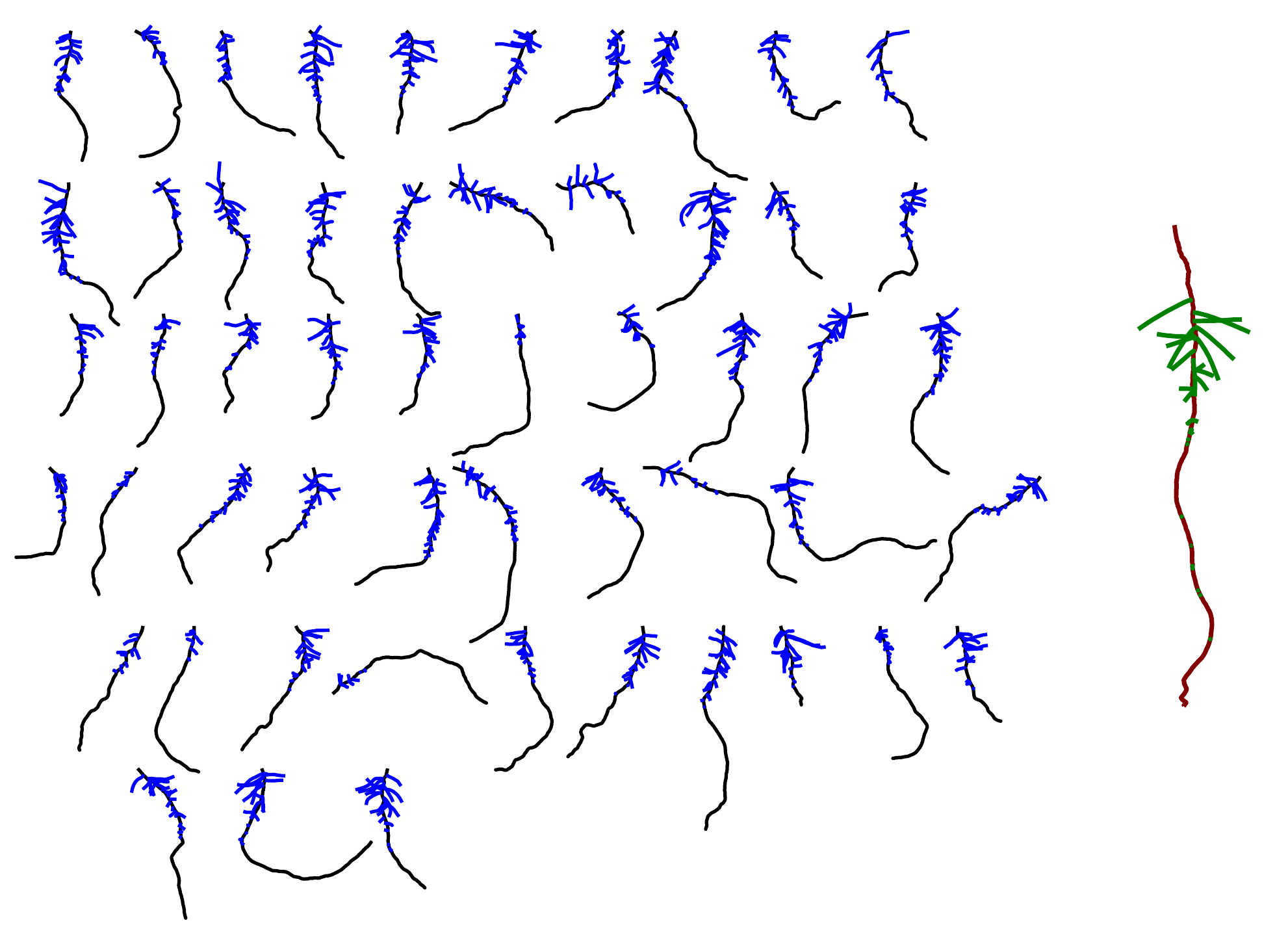}
	\caption{Example of a mean root  computed using the proposed approach. The input collection of 53 roots is shown on the left and the computed mean root is shown on the right with different colors. The mean root has been enlarged for visual clarity.}
	\label{fig:meanRootForRootCollection}
\end{figure*}

\subsection{Summary statistics}

Figure~\ref{fig:meanRootForRootCollection} shows the mean root computed from the entire data set. The 53 input roots are shown on the left while the generated mean root is shown on the right. Note that the correspondences between each pair in the dataset, as well as between each sample in the dataset and the computed mean root, are automatically computed and do not require any user input or interaction. Also, the average of a pair of plant roots is a by-product of the geodesic computation process. In fact, the root that is equidistant to the source and the target roots is exactly  their mean. Thus, the root that lies exactly at the middle of each geodesic in Figure~\ref{fig:geodesicComparison_1} is actually the average of the leftmost and rightmost roots.

\begin{figure*}[t]
\center
\includegraphics[width=0.8\textwidth]{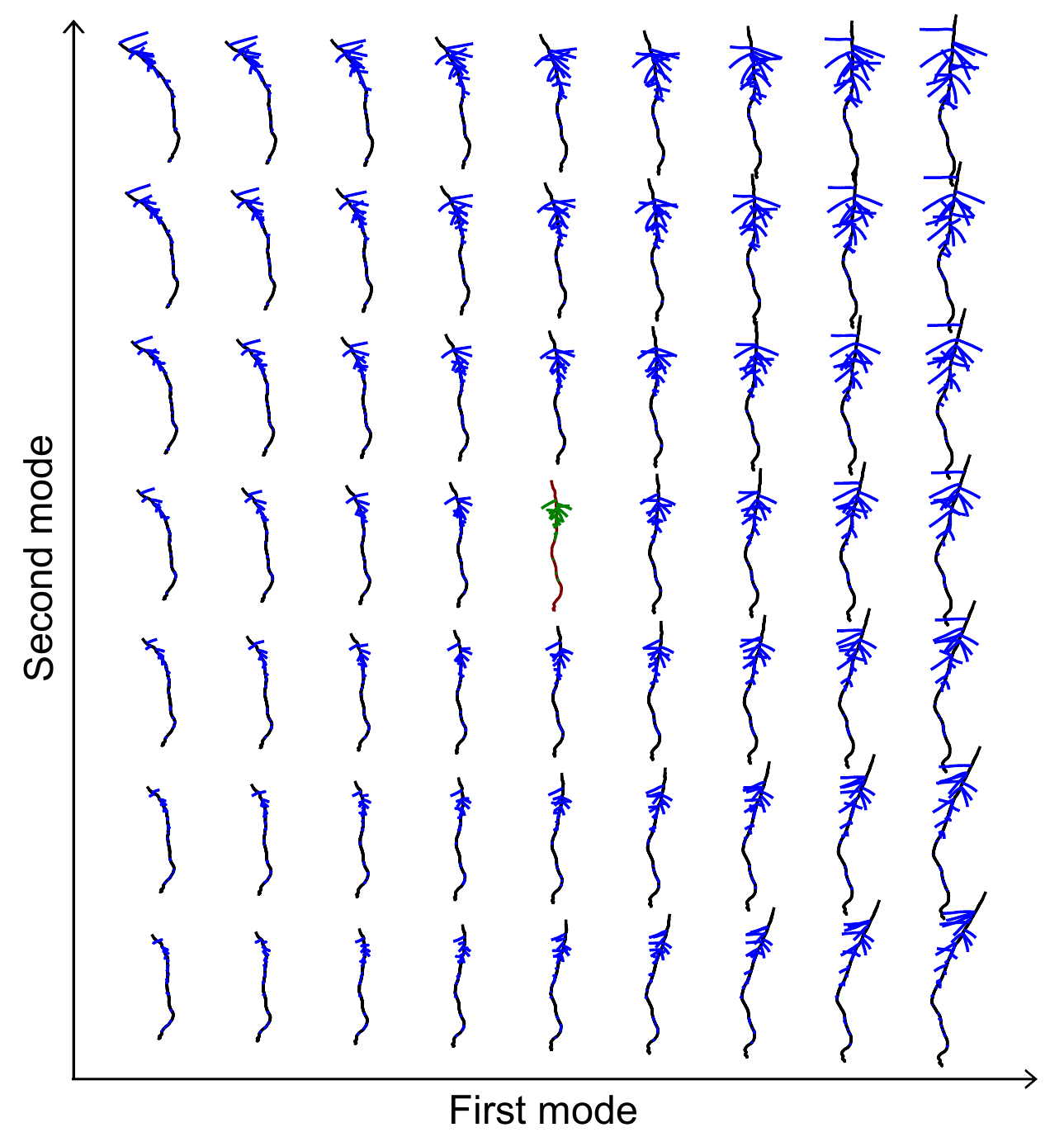}
\caption{The first two modes of variations computed on the dataset of  Figure~\ref{fig:meanRootForRootCollection}.  The mean root in the centre of the array is highlighted using the same colors as  of Figure~\ref{fig:meanRootForRootCollection}.}
\label{fig:principalVariationModes}
\end{figure*}

Figure~\ref{fig:principalVariationModes} shows the first two principal modes of variation for the root dataset of Figure~\ref{fig:meanRootForRootCollection}. From these results, we clearly see that the first two leading modes capture the main geometric and topological variations in the root dataset.

\subsection{Random root synthesis}
\begin{figure*}[t]
\center
\includegraphics[width=1.0\textwidth]{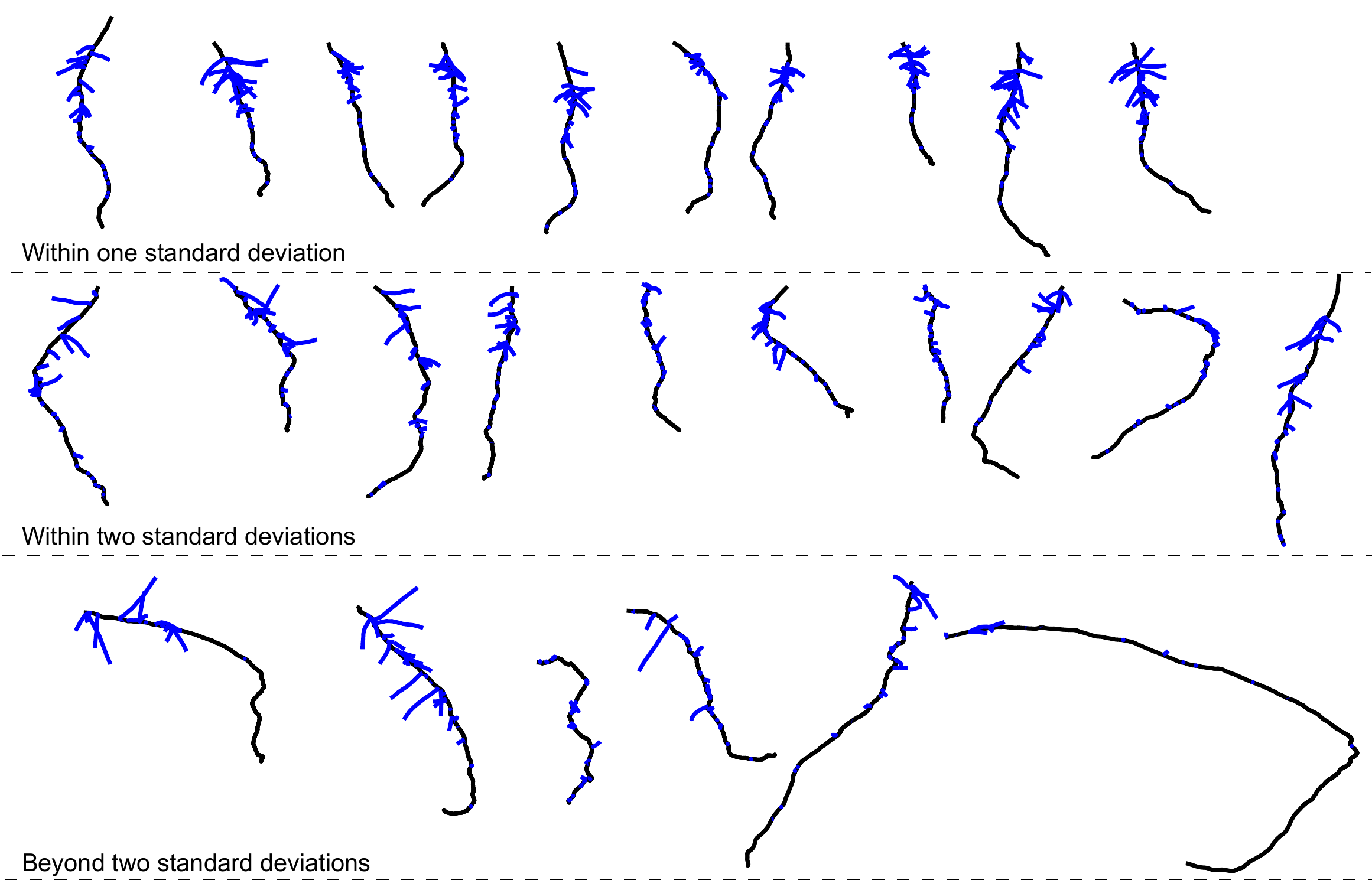}
\caption{Root shapes randomly sampled from the Gaussian  distribution fitted to the  root dataset of Figure~\ref{fig:meanRootForRootCollection}.}
\label{fig:randomRoots}
\end{figure*}

Figure~\ref{fig:randomRoots} shows plant roots that have been automatically synthesized by random sampling from the statistical model fitted to  the root dataset of Figure~\ref{fig:meanRootForRootCollection}.  These roots have been synthesized without any biological knowledge or interaction from the users.  For a better visualization, we group these synthesized plant roots into three clusters depending on their distances to the mean root.  Note that those that are close to the mean share some similarities with the input roots but are not the same. The further the synthesized roots deviate from the mean the less plausible they become.

\subsection{User-controlled root synthesis}

\begin{figure*}[t]
\center
\includegraphics[width=1.0\textwidth]{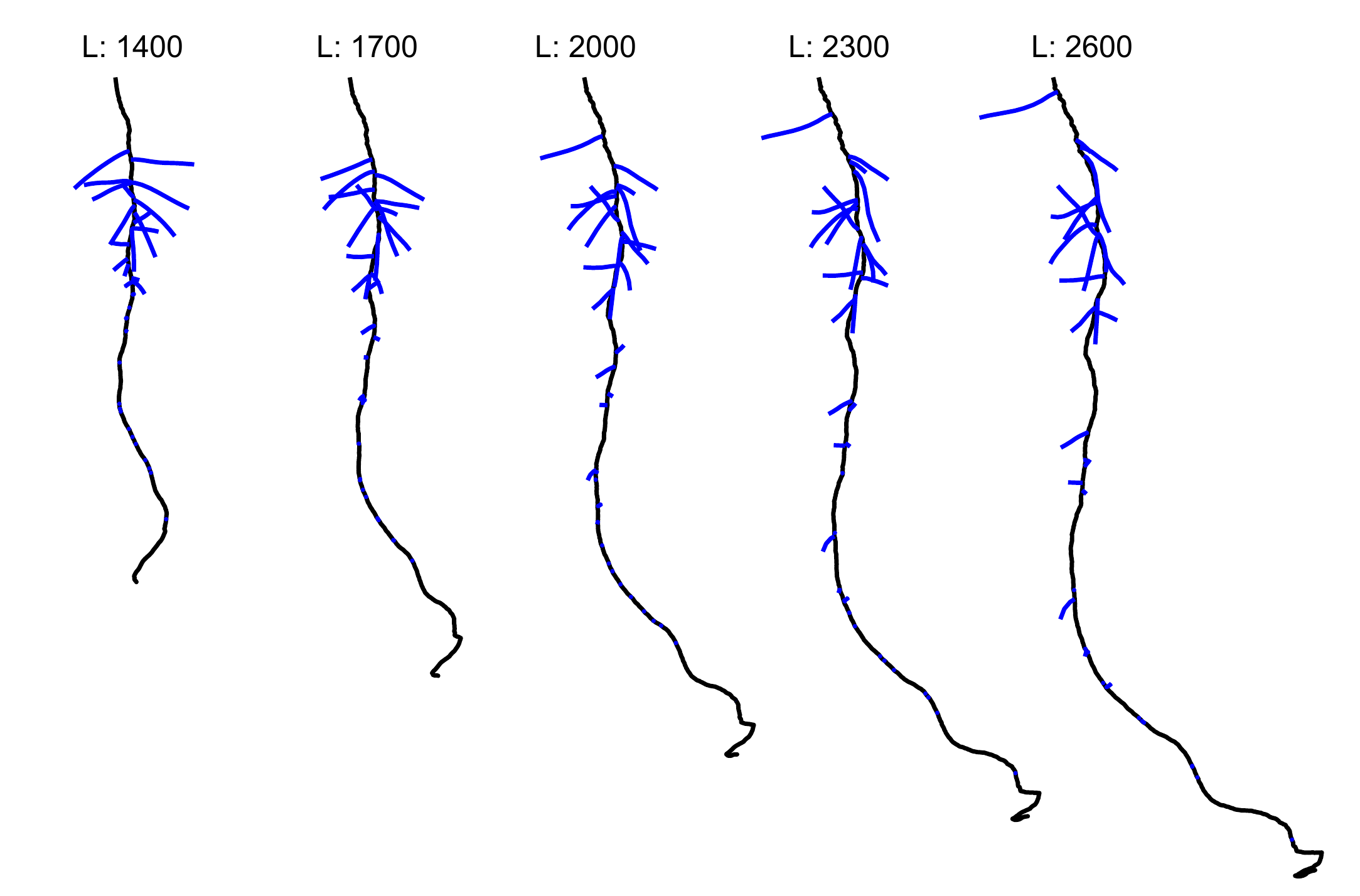}
\caption{Root shapes generated by user-specified control parameters. In this example, the user specifies the length of the main root, and the system automatically synthesizes root shapes.}
\label{fig:regression_length}
\end{figure*}

\begin{figure*}[t]
\center
\includegraphics[width=0.8\textwidth]{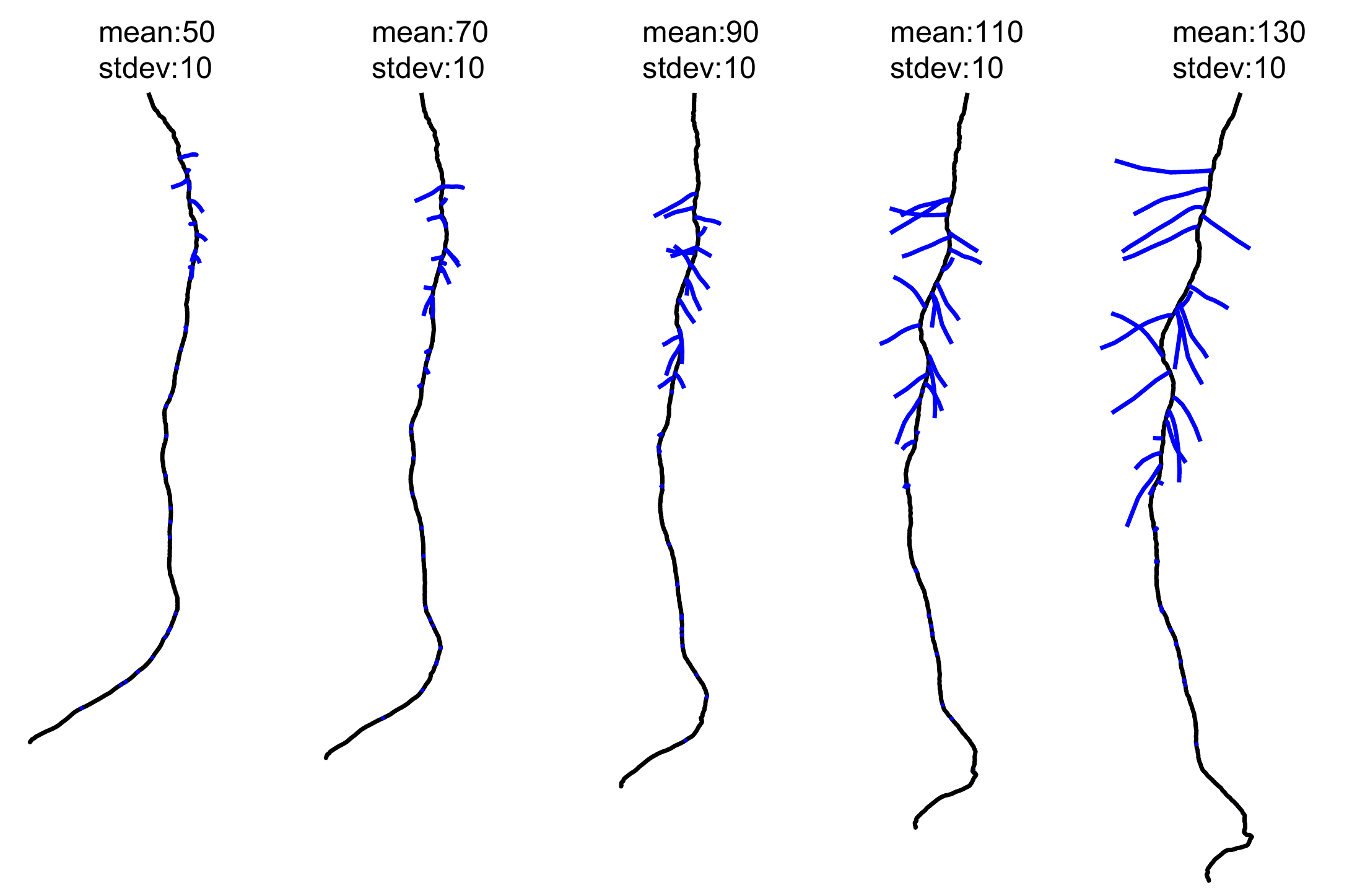}
\caption{We consider the mean and standard deviation of the lengths of the lateral branches.  The root shapes have been generated by fixing the standard deviation of the length of the lateral  roots and varying the value of the mean of the lateral  root lengths.}
\label{fig:regression_sideRootLen-1}
\end{figure*}

\begin{figure*}[t]
\center
\includegraphics[width=0.8\textwidth]{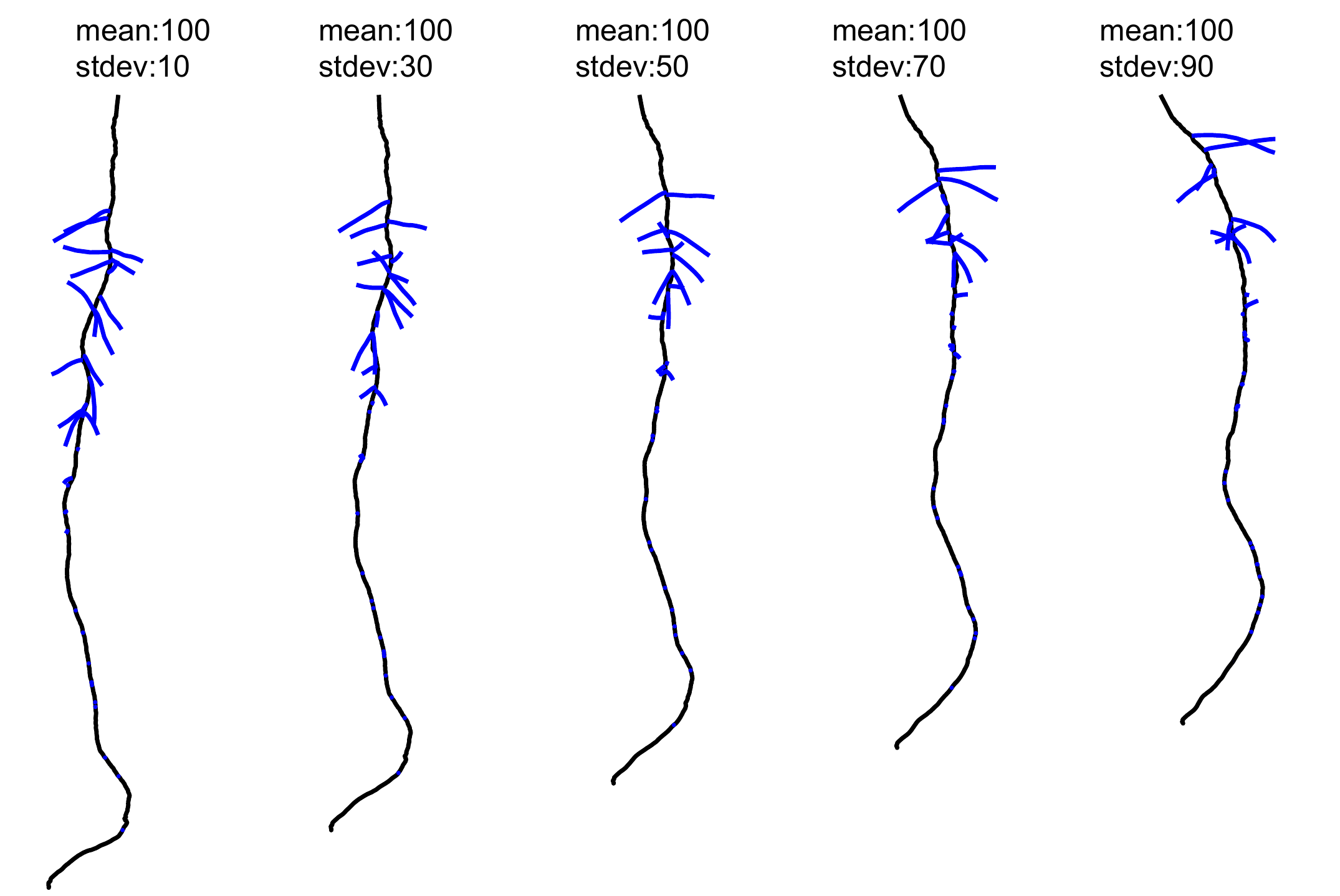}
\caption{We consider the mean and standard deviation of the lengths of the lateral  roots. The figure shows root shapes generated by fixing the value of the mean of the lateral root lengths  and varying the standard deviation of the length of the lateral roots. }
\label{fig:regression_sideRootLen-2}
\end{figure*}

We show a few examples of plant roots generated with intuitive controls such as \emph{main root length}, \emph{mean length of side-roots}, and \emph{standard deviation of side roots length}. We take the plant roots of Figure~\ref{fig:meanRootForRootCollection}, compute their mean root and their modes of variation, and then learn a regression model that maps the three control parameters into points in the tree-shape space. This  provides a direct way to explore the range of root shapes. Figure~\ref{fig:regression_length},~\ref{fig:regression_sideRootLen-1},~\ref{fig:regression_sideRootLen-2} show a few representative results produced with this approach.

\subsection{Timing}

The propsoed framework has been implemented in Matlab and runs on a desktop PC with Intel(R) Core(TM) i5-4460 CPU@3.20GHz and 8GB of RAM. Table~\ref{tab:timing_geodesics} reports the computation time for each geodesic shown in Figure~\ref{fig:geodesicComparison_1}. It also reports the side root number of the roots after adding virtual side roots.

To compute each of the geodesics of Figure~\ref{fig:geodesicComparison_1}, we first align the source and target roots to one another (alignment step), and then compute their geodesic using the method proposed in this paper, the method of Feragen \etal~\cite{feragen2013toward}, and the method of Wang \etal~\cite{wang2018the}.  Table~\ref{tab:timing_geodesics} shows that most of the time is consumed in the alignment process. Once the roots are aligned, the method proposed in this article for computing geodesics  is significantly faster  than the approaches of Feragen \etal \ and Wang \etal  \ ($40$ms for ours vs. $11$s for the other two methods).

The computation of the mean root takes $9$ hours $4$ minutes. Once the mean root has been computed, computing the principal modes of  variation  takes $1.1$ seconds. It only takes  $0.22$ seconds to generate one random root. For regression, it only tales  $0.01$ seconds to compute the mapping matrix $M$ and about $0.001$ seconds (on average) to generate a new root. This suggests that the approach proposed in  this article is well suited for the generation and synthesis of root structures

\begin{table*}[t]
\small
\center{
\caption{\label{tab:timing_geodesics} Comparison of the computation time (in seconds) between the proposed approach and the approaches of Feragen \etal~\cite{feragen2013toward}  and Wang \etal~\cite{wang2018the}. "Size" refers to the number of lateral roots. }
\resizebox{\textwidth}{!}{
\begin{tabular}{@{}ccccccc@{}}
\hline
		 &  Size &  Approach &  Alignment & Correspondence & Geodesic & Overall \\
\hline
Fig.~\ref{fig:geodesicComparison_1} - Example 1&  $23$ & This paper&  $147.0$ & N/A & $0.04$ &  \textbf{$147.04$}\\

    &  $23$ &  Feragen \etal~\cite{feragen2013toward}   & $147.0$ & N/A & $10.98$ & $157.98$ \\

    & $23$ & Wang \etal~\cite{wang2018the}  & $147.0$ & $0.0050$ & $11.40$ & $158.40$\\
\hline
Fig.~\ref{fig:geodesicComparison_1} - Example 2       & $14$  & This paper  &  $137.0$ & N/A & $0.04$ & \textbf{$137.04$} \\

    & $14$ & Feragen \etal~\cite{feragen2013toward}  & $137.0$ & N/A & $10.60$ &  $147.60$ \\

    & $14$ &  Wang \etal~\cite{wang2018the}              & $137.0$ & $0.0028$ & $11.31$ & $148.31$ \\

\hline
\end{tabular}}
}
\end{table*}

\subsection{Application to plant root classification}

\begin{figure*}[t]
\center
\includegraphics[width=1.0\textwidth]{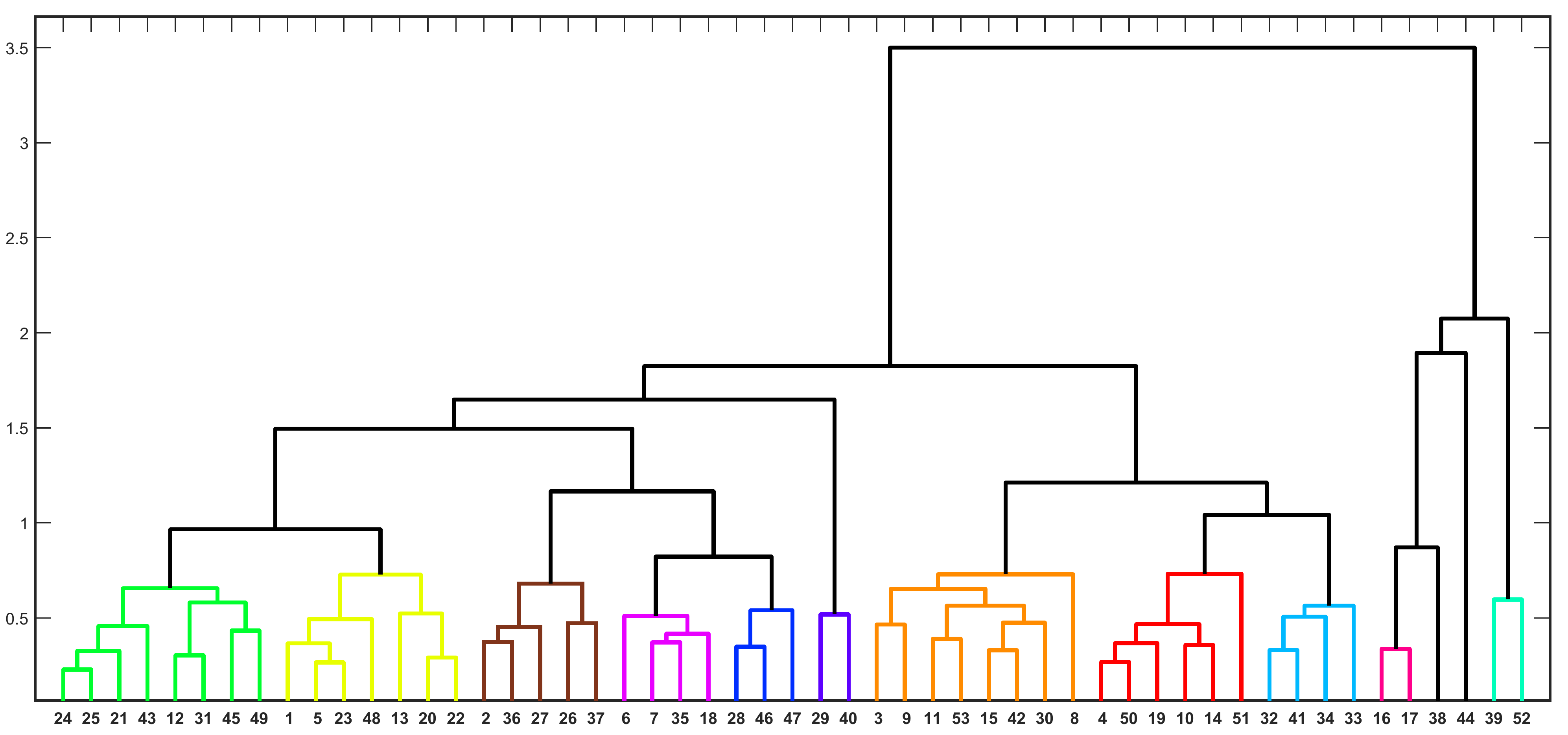}
\caption{The hierarchical binary clustering result of the roots in Figure~\ref{fig:Classification-clusters}.}
\label{fig:Classification-hierarchical_clustering}
\end{figure*}

\begin{figure*}[t]
\center
\includegraphics[width=1.0\textwidth]{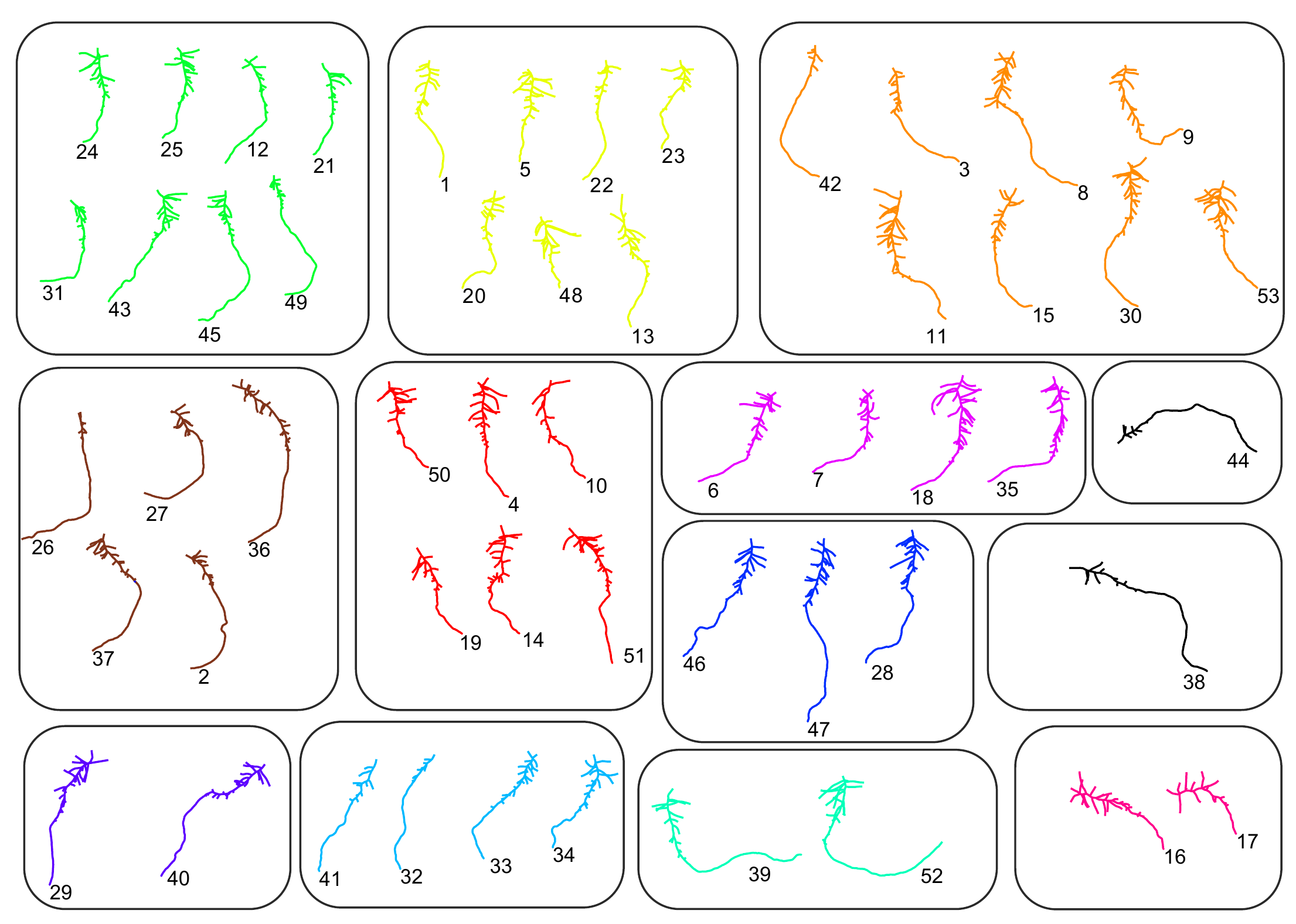}
\caption{The clustering results for input roots in accordance with the results of Figure~\ref{fig:Classification-hierarchical_clustering}. Roots within each cluster are drawn using the same color.}
\label{fig:Classification-clusters}
\end{figure*}

Finally, the length of the geodesic between two roots is a measure of dissimilarity between these roots. To demonstrate the utility of this measure for plant root classification, we computed the pairwise geodesic distance matrix for the entire dataset and performed a hierarchical binary clustering using MATLAB's linkage function. Figure~\ref{fig:Classification-clusters} shows the root clustering result according to the hierarchical binary clustering result as is shown in Figure~\ref{fig:Classification-hierarchical_clustering}. As one can see, the metric clusters together roots that have similar geometry and topology.

\section{Conclusion and Future Work}
\label{sec:conclusion}
In this article, we have developed a full framework for the statistical analysis and modeling of the geometry and topology of plant roots. Central to the framework is an elastic curve-based tree-shape space and an elastic  metric that enables us to find optimal correspondences between roots, even in the presence of large topological deformations. As shown in the experimental results, the metric is suitable for comparing roots based on their shape,  and for computing geodesics between them. We showed that the representation and metric can be used to compute statistical summaries of root collections and to synthesize new roots either randomly or interactively by varying biologically-motivated  parameters. This can lead to many practical applications. For instance, the proposed framework can be used alongside with root imaging techniques, which are recently receiving a growing attention, see for example~\cite{rootdata2019} for a repository of root datasets. The proposed geodesic computation algorithm  and geodesic distance  can be used to compare Root System Architectures (RSA) and to understand and model similarities and differences of RSAs within and across plant species. The tools for computing statistics (means and modes of variations) can be used to model the variability in geometry and topology of RSAs within species. It can also be used for probabilistic classification. Finally,  regression tools are fundamental to the simulation and modelling of  RSAs from biological or environmental parameters. To the best of our knowledge, this is the first paper that provides all the building blocks for these high-level tasks.

Although effective as evidenced by the results shown in this article, there are still some limitations that can be addressed in future work. First, we only considered roots that have two layers of branches. However, roots have complex tree structures  often composed of more than two layers. In the future, we plan to extend the current framework to processing multi-layer roots. Second, the current formulation does not  take into account the thickness of the roots. As such, the framework does not differentiate between coarse and fine roots. While root thickness can be easily taken into account by adding a fourth term to Equation~\eqref{equ:init_distanceTwoRoots}, it will result in an additional parameter to tune and an increase in computation time. As such, we plan in the future to explore more efficient techniques.  Third,  the type of geodesics that our framework generates between a pair of roots depends on the weights of the  three terms of the metric defined in Equation~\eqref{equ:init_distanceTwoRoots}.  While in this article, we manually set these weights, in practice they depend on the application and on the plant species being analysed. As such, it would be interesting to learn these parameters from data.  Third, we have demonstrated that this framework can be  used to regress plant root shapes  from a few parameters. Our framework is general and can be easily extended to  incorporate  biological knowledge and environmental effects such as soil humidity and nutrient content. In the future, it will be interesting to incorporate these factors into the regression process and explore more regression tools beyond the linear ones.

Finally, we plan in the future to use the proposed  framework to quantify differences in 3D root morphology, and  comparing 3D root systems of either different genotypes grown under the same environmental conditions, or 3D root systems of the same genotype plant grown under different environmental conditions, including different nutrient concentrations, or soils at different levels of moisture content, or soils of different toxicity. This could be achieved either simply by calculating the geodesic distance between mean roots, or by using advanced statistical measures that capture variability within each group of plants. 


\vspace{6pt}
\noi\textbf{Acknowledgement. } The authors would like to thank Adam Duncan for sharing the code for analyzing simple neuronal trees, and Aasa Feragen for making the code for tree-like shape analysis publicly available. Guan Wang would like to thank the China Scholarship Council for funding his visit and stay at Murdoch University, Australia.  

\bibliographystyle{IEEEtran}
\bibliography{references_v2}

\end{document}